\documentclass[%
 reprint,
superscriptaddress,
showkeys,
nofootinbib,
 amsmath,amssymb,
 aps,
floatfix,
]{revtex4-1}
\usepackage{accents}
\usepackage{graphicx}
\usepackage{dcolumn}
\usepackage{bm}
\usepackage[utf8]{inputenc}
\usepackage{tabularx}
\usepackage[dvipsnames]{xcolor}
\usepackage{units}
\usepackage[euler]{textgreek}
\usepackage{upgreek}
\usepackage{derivative}
\usepackage{subcaption}
\usepackage{amsmath}
\usepackage{physics}
\usepackage{comment}
\usepackage{amssymb}
\usepackage[justification=justified,singlelinecheck=false]{caption}
\usepackage[colorlinks=true,linkcolor=blue!50!black,urlcolor=blue!50!black,citecolor=blue!50!black]{hyperref}
\usepackage{ragged2e}

\graphicspath{ {img/} }

\begin{document}

\preprint{APS}

\title{Sub-GeV Dark Matter Under Pressure from Direct Detection}
\author{Andrew Cheek}
\email{acheek@sjtu.edu.cn}
\affiliation{Tsung-Dao Lee Institute \& School of Physics and Astronomy, Shanghai Jiao Tong University, Shanghai 201210, China}
\author{Pablo Figueroa}
\email{pfigueroaf@unal.edu.co}
\affiliation{Instituto de Física Corpuscular (IFIC), Consejo Superior de Investigaciones Científicas (CSIC) \\
 and Universitat de València, C/ Catedratico Jose Beltran 2, E-46980 Paterna, Spain}

\author{Gonzalo Herrera}
\email{gonzaloherrera@vt.edu}
\affiliation{Center for Neutrino Physics, Department of Physics, Virginia Tech, Blacksburg, Virginia 24061, USA}

\author{Ian M. Shoemaker}
\email{shoemaker@vt.edu}
\affiliation{Center for Neutrino Physics, Department of Physics, Virginia Tech, Blacksburg, Virginia 24061, USA}

\begin{abstract}
The DAMIC-M collaboration recently reported impressive bounds on sub-GeV dark matter, robustly testing both thermal and non-thermal models for the very first time. In this work we derive novel bounds from the recent PandaX-4T ionization S2-only search for Coherent Elastic Neutrino-Nucleus Scattering (CE$\nu$NS). We find that the PandaX-4T S2-only data is able to compete with the DAMIC-M results, providing the best constraints for scalar and asymmetric thermal dark matter models for masses between 20 to 200 MeV. We further discuss the implications of recent direct detection results for several other sub-GeV dark matter models, highlighting their complementarity with astrophysical, cosmological and laboratory probes.

\end{abstract}

\maketitle

\section{Introduction}

As a consequence of direct dark matter detection experiments reaching impressive sensitivities but not observing a dark matter signal, the last decade and a half has seen the community diversify its search strategy away from GeV-TeV scale Weakly Interacting Massive Particles (WIMPs). One of the most promising new directions experimentally has been focused on the direct detection of MeV scale dark matter \cite{Essig_2012,Essig:2012yx,Catena:2019gfa,akerib2022snowmass2021,XENON:2019gfn,CRESST:2019jnq,XENON:2024znc, PandaX-II:2021nsg, PandaX:2022xqx, SENSEI:2020dpa, SENSEI:2023zdf, SuperCDMS:2024yiv, Balan:2024cmq}. The initial theoretical sparks for such dark matter models were already present in the literature. In particular, it was shown that the standard freeze-out scenario could accommodate MeV scale dark matter for a number of particle physics models~\cite{Boehm:2003hm,Boehm:2003ha, Pospelov:2007mp, Hooper:2008im}. Beyond this, more possibilities have been explored, such as asymmetric dark matter \cite{Kaplan:2009ag, Graesser:2011wi, Lin:2011gj}, freeze-in \cite{Hall:2009bx,Essig_2012, Elor:2021swj, Bhattiprolu:2023akk, Boddy:2024vgt}, as well more complex scenarios involving decoupling via scattering processes or $3\rightarrow 2$ processes in the dark sector \cite{Hochberg:2014dra,Kuflik:2015isi, DAgnolo:2017dbv, Smirnov:2020zwf, Chu:2024rrv}.

The simplest of these scenarios remained widely untested until very recently. The results from the DAMIC-M collaboration changed this, ruling out large portions of parameter space of some of these models \cite{DAMIC-M:2025luv,Krnjaic:2025noj}. Here we derive novel bounds from the recent CE$\nu$NS search in the S2-channel at PandaX-4T, finding that this rules out new regions of parameter space where the relic abundance can be obtained in these models. The recent results from DAMIC-M and PandaX-4T S2 channel search for CE$\nu$NS put significant pressure on sub-GeV dark matter, and deserve a dedicated analysis and contextualization in the landscape of different models and complementary probes.

The paper is organized as follows: In Section \ref{sec:rate}, we describe the formalism used to compute the dark matter-electron ionization rate at PandaX-4T S2 channel, derive limits on the dark matter-electron scattering parameters. In Section \ref{sec:models}, we contextualize our results on the landscape of sub-GeV dark matter models able to yield the observed relic abundance. In Section \ref{sec:complementary}, we discuss complementary bounds to direct detection of galactically bounded dark matter, both with direct detection and with complementary probes. Finally, in Section \ref{sec:discussion}, we present our conclusions. The paper is accompanied by several appendices with details on our treatment of PandaX-4T data.

\begin{figure*}[t] 
\centering 
\includegraphics[scale=0.58]
{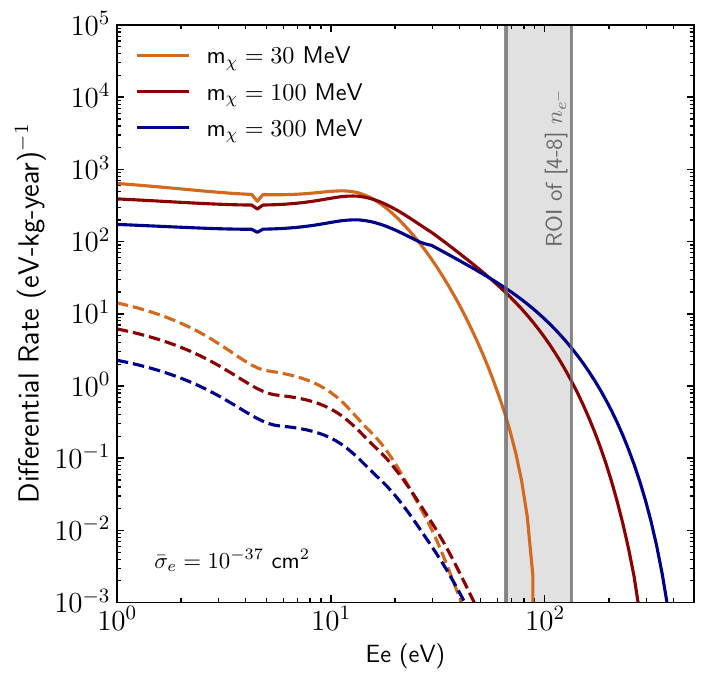}
\hspace{5mm}
\includegraphics[scale=0.49]{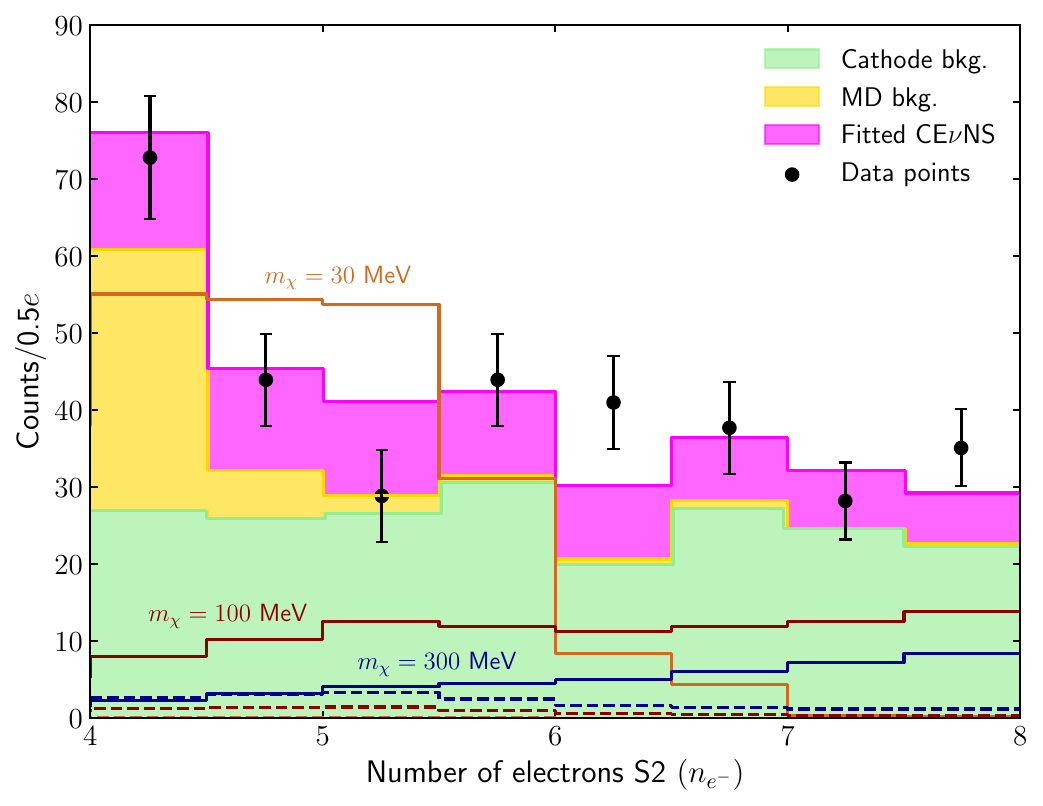}
\caption{\justifying \textit{Left panel}: Differential ionization rate in liquid xenon as a function of the electron recoil energy $E_{e}$, for three benchmark values of the dark matter mass, considering a heavy (solid) and ultralight (dashed) mediator. For comparison, we show the ROI of PandaX-4T S2-only search for CE$\nu$NS as a shaded browm band. \textit{Right panel:} Counts on PandaX-4T S2 channel in terms of $n_{e}$ for the Lindhard model, matching the dark matter-induced electron recoil rate with the binning of the data and background given in~\cite{pandaxcollaboration2024indicationsolar8bneutrino}. The ROI is indicated as the shaded grey region. 
}
\label{fig:counts-PandaX-DMe}
\end{figure*}
\section{Dark matter-electron scattering at PandaX-4T}\label{sec:rate}

Dark matter-electron scatterings at liquid xenon detectors induce a differential event rate in electron ionization energy $E_e$

\begin{equation}\label{eq:dRdlnEe}
    \frac{dR}{d \ln{E_{e}}} = N_{T} \sum_{n,l} \mathcal{A}(E_{e}) \int d^{3}v \ \mathcal{F}(\vec{v} + \vec{v_{e}}) \frac{d \sigma (v,E_{e})}{d \ln{E_{e}}},
\end{equation}
with $N_T$ denoting the number of targets in the detector, and with efficiency function $\mathcal{A}(E_{e})$. This function depends on the quenching factor model considered. We present a discussion on this and related different choices in Appendix \ref{app:quenching}.

The differential ionization cross section is given by 

\begin{equation}
    \frac{d \sigma (v,E_{e})}{d \ln{E_{e}}} = \frac{\bar{\sigma}_{e}}{8 \mu^{2}_{\chi e}v^{2}} \int_{q_{\rm min}^{nl}}^{q_{\rm max}^{nl}}dq \,q \, |f_{\rm ion}^{nl}(\vec{k},q)|^{2} |F_{\rm DM}(q)|^{2},
\end{equation}
where $\vec{k}$ denotes the atomic electron momentum and $q$ denotes the momentum transfer of the dark matter-electron scattering process. The ionization form factor reads
\begin{equation}
f_{\rm ion}(\vec{k}, q)=\frac{2|\vec{k}|^3}{(2 \pi)^3} \int \frac{\mathrm{~d}^3 \vec{k}}{(2 \pi)^3} \psi_{k^{\prime} \ell^{\prime} m^{\prime}}^{\prime}(\vec{k}+\vec{q}) \psi_{n l m}(\vec{k}).
\end{equation}

We use the tabulated wavefunctions $\psi_{klm}$ for xenon from \texttt{QEDark} \cite{Essig:2015cda, jelle_aalbers_2023_7636982}. The event rate from Eq.~\ref{eq:dRdlnEe} contains an integral over velocity of the dark matter particle and momentum transfer of the scattering processes. The minimum dark matter velocity needed to ionize an electron in the $(n, l)$ shell with outgoing energy $E_{e}$ is given by

\begin{equation}
v_{\min }=\frac{E_{e}+\left|E^{n l}\right|}{|\vec{q}|}+\frac{|\vec{q}|}{2 m_\chi},
\end{equation}
where $E^{n l}$ is the binding energy of the atomic electron. The integration over momentum transfer is performed in the range

\begin{equation}
\begin{aligned}
& q_{\min }=m_\chi v-\sqrt{m_\chi^2 v^2-2 m_\chi\left(E_{e}+\left|E^{n l}\right|\right)} \\
& q_{\max }=m_\chi v+\sqrt{m_\chi^2 v^2-2 m_\chi\left(E_{e}+\left|E^{n l}\right|\right)}.
\end{aligned}
\end{equation}

$\mathcal{F}(\vec{v} + \vec{v_{e}})$ denotes the dark matter flux reaching the detector, and reads
\begin{equation}
\mathcal{F}\left(\vec{v}+\vec{v}_E\right)=\frac{\rho_\chi}{m_\chi} f\left(\vec{v}+\vec{v}_E\right),
\end{equation}

where $\rho_{\chi}=0.4$ GeV/cm$^3$ is the local dark matter density in the solar system \cite{Read:2014qva}, and $f(\vec{v}+\vec{v}_{E})$ is the dark matter velocity distribution function, where $\vec{v}$ is the dark matter velocity in the rest frame of the detector, and $\vec{v}_E$ is the velocity of the Earth with respect to the Galactic frame. We use the Standard Halo Model for our benchmark calculations, corresponding to a Maxwell-Boltzmann distribution truncated at the escape velocity of the Milky Way \cite{Herrera_2021,Green:2011bv}. In Section \ref{sec:complementary}, we discuss the impact of uncertainties of the velocity distribution on our results. The momentum-dependence of the scattering process is model-dependent and it is typically encoded in the DM form factor, defined as
\begin{equation}
F_{\rm DM}(q) = \frac{(m_{A^{\prime}}^{2} + \alpha^{2}m_{e}^{2})}{(m_{A^{\prime}}^{2} + q^{2})},
\end{equation}
where $m_{A^{\prime}}$ denotes the mediator mass. Defining this form factor allows to factorize from the ionization rate a non-relativistic dark matter-electron scattering cross section at fixed momentum transfer $q_e=\alpha m_e$

\begin{equation}
\bar{\sigma}_e=\frac{16 \pi \alpha \alpha_D \epsilon^2 \mu_{\chi e}^2}{\left(m_{A^{\prime}}^2+\alpha^2 m_e^2\right)^2}
\end{equation}
where $\alpha_{\chi}=g_{\chi}^2/4\pi$ parametrizes the coupling strength of the dark matter to the new mediator $\phi$. We provide more concrete details for different models in Section \ref{sec:models}.

To compute the expected number of events in each bin $i$ of PandaX-4T, and to allow for direct comparison with the binned data reported in Ref. \cite{pandaxcollaboration2024indicationsolar8bneutrino}, we perform the conversion from electron ionization energy $E_e$ to number of observable electrons $n_{e^{-}}$ described in \cite{Essig:2012yx,Essig_2017} as
\begin{equation}\label{eq:conversion_ne}
E_e=\left(n_\gamma+n_{e^{-}}\right) W,
\end{equation}
where $n_{\gamma}$ is the number of scintillation photons, and $W=13.8$ eV is the energy required to produce a quanta of energy in electrons or photons. Further details on the conversion from $E_{e}$ to $n_e^{-}$ are provided in Appendix~\ref{app:quenching}. Then, we compute the total rate in each bin $i$ as

\begin{equation}\label{Ntot}
    s_i(\bar{\sigma}_{e},m_{\chi}) = \mathcal{E} \int_{n_{e^{-}}^{\rm min,i}}^{n_{e^{-}}^{\rm max,i}} \frac{dR}{dn_{e^{-}}} dn_{e^{-}}
\end{equation}

where $\mathcal{E}$ denotes the exposure of the experiment. For the PandaX-4T S2-only CE$\nu$NS search in \cite{pandaxcollaboration2024indicationsolar8bneutrino}, this corresponds to $\mathcal{E}=1.04$ tonne$\cdot$yr. The Region of Interest (ROI) in number of ionized electrons is  $n_{e^{-}}$ = [4 - 8], corresponding to $E_{e} =$ [66 - 133] eV. This ROI is separated in 8 bins. For illustration, we show the differential event rate in terms of electron recoil energy for different dark matter masses in the left panel of Fig. \ref{fig:counts-PandaX-DMe}, both for a heavy mediator $F_{\rm DM}=1$ (solid lines) and for a massless mediator $F_{\rm DM}=\alpha^2m_{e}^2/q^{2}$ (dotted lines). We also show the  ROI of PandaX-4T as a band. From this Figure it is noticeable that dark matter masses below $m_{\chi} \lesssim 20$ MeV are kinematically inaccesible at PandaX-4T.

\begin{table}[h]
    \centering
    \begin{tabular}{|c|c|c|}
     \hline
     PandaX-4T S2 Only & Run0 & Run1 \\
     \hline
     \hline
     Cathode & 100 $\pm$ 24 & 104 $\pm$ 21 \\
     MD & 25 $\pm$ 3 & 20 $\pm$ 4 \\
     ERs & 1.3 $\pm$ 0.1 & 0.9 $\pm$ 0.2 \\
     \hline
     Total bkg. & 126 $\pm$ 24 & 125 $\pm$ 21 \\ 
     $^{8}$B CE$\nu$ES & 18 $\pm$ 4 & 25 $\pm$ 6 \\
     \hline
     \hline
     Expected & 144 $\pm$ 25 & 150 $\pm$ 22 \\ 
     \hline
     Observed & 158 & 174 \\
     \hline
     \hline
     \multicolumn{3}{|c|}{Combined (Run0 + Run1)} \\ 
     \hline
     \multicolumn{3}{|c|}{
            \begin{tabular}{c|c}
                Expected \hspace{5mm} & \hspace{5mm} 294 $\pm$ 33 \\
                Observed \hspace{5mm} & \hspace{5mm} 332 \\
            \end{tabular}
        } \\ \hline
     \end{tabular}
     \caption{\justifying PandaX-4T S2 channel event rate expectations and observations in their ROI for the CE$\nu$NS search.}
     \label{tab:recoilexcess}
\end{table}

For clarity, we show the total background, expectations and data reported by PandaX-4T within the ROI of their S2-analysis, see Table ~\ref{tab:recoilexcess}. We further attempted to independently reproduce the CE$\nu$NS expectations in each bin, see the discussion in Appendix \ref{app:sevens}. 

\begin{figure*}[t] 
    \vskip\baselineskip
    \begin{subfigure}{0.49\textwidth}
        \centering
        \includegraphics[scale=0.55]{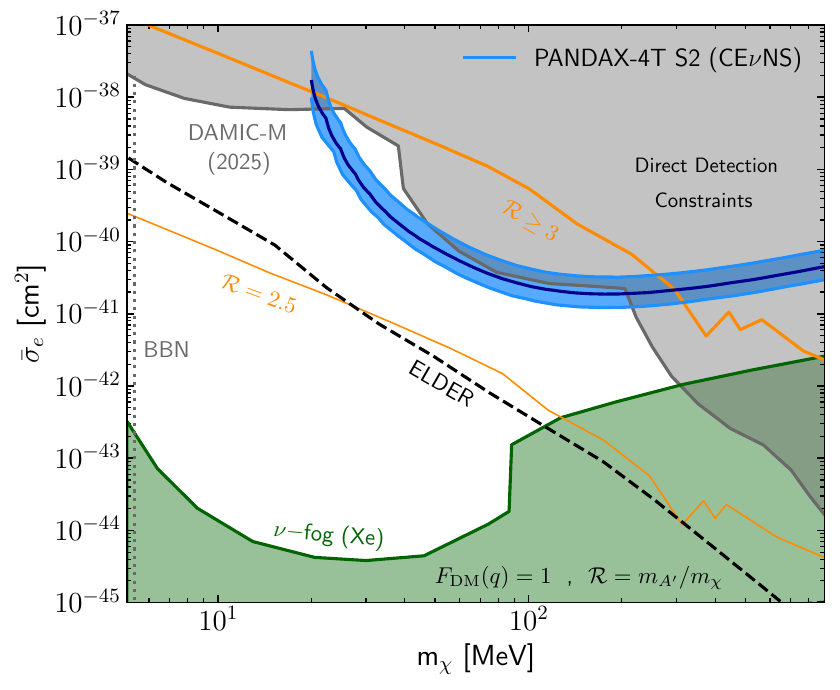}
        
    \end{subfigure}
    \hfill
    \begin{subfigure}{0.49\textwidth}
        \centering
        \includegraphics[scale=0.55]{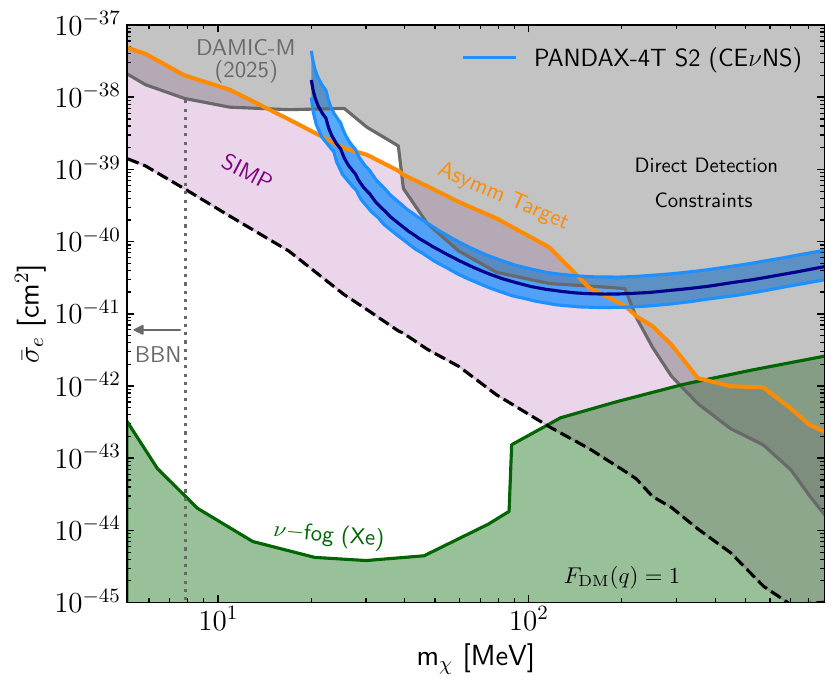}
        
    \end{subfigure}
     \caption{\justifying \textit{Left panel:} Upper limits on the non-relativistic dark matter-electron scattering cross section $\bar{\sigma}_{e}$, from PandaX-4T S2-channel search for CE$\nu$NS. The limits are displayed as a band, to account for different quenching factor models, see Appendix \ref{app:quenching}. The solid blue line corresponds to the best-fit quenching model at PandaX-4T. For the heavy mediator case, for comparison, we show in a shaded grey region the previous bounds obtained for other dark matter direct detection searches at DAMIC-M \cite{DAMIC-M:2025luv}, PandaX \cite{PandaX:2022xqx}, XENONnT, and CRESST-III \cite{CRESST:2019jnq}. We also show in grey color a vertical line indicating the lower bound on the dark matter mass obtained from measurements of $N_{\rm eff}$ at BBN and CMB \cite{Giovanetti:2021izc}. We further show in shaded green color the neutrino fog arising from neutrino-electron elastic scatterings at the detector \cite{Carew:2023qrj, Wyenberg:2018eyv}. We show in orange lines benchmark regions of parameter space where scalar dark matter can induce the observed relic abundance of dark matter in the Universe \cite{Krnjaic:2025noj}, for different ratios of the dark matter to mediator mass. We further show as a shaded purple band the region where SIMP/ELDER can yield the relic abundance \cite{Hochberg:2014dra, Kuflik:2015isi}. Further on the different models are provided in Section \ref{sec:models}.  \textit{Right panel:} Limits on $\bar{\sigma_e}$ for asymmetric dark matter, confronted with the smallest possible relic abundance target in orange color \cite{Lin:2011gj}. Smaller cross sections are ruled out by energy injection bounds on the CMB.}
    \label{fig:x-section-results}
\end{figure*}

As previously mentioned, to derive a stringent upper limit, we make use of the spectral information in the ROI of PandaX-4T \cite{pandaxcollaboration2024indicationsolar8bneutrino}. In Figure \ref{fig:counts-PandaX-DMe}, we show the number of counts in terms of number of electrons for the S2-channel CE$\nu$NS search in PandaX-4T. The black data points show the observed events in each bin. The green histograms show the expected events from radioactivity in the cathode electrode, the orange histogram shows the expected events from the micro-discharging background, and the purple histograms show the expected events from CE$\nu$NS. We performed a dedicated discussion of the expected CE$\nu$NS events in the Appendix \ref{app:sevens}, finding small but sizable discrepancies in some of the energy bins. We predict a larger number of events, which would lead to a tighter limits on the dark matter-electron scattering parameters, however, we conservatively use the benchmark expectations reported by the collaboration in our analysis.
We further show in Figure \ref{fig:counts-PandaX-DMe} the expected counts induced by dark matter-electron scattering, via a heavy mediator, and for different values of the dark matter mass. It can be noticed that the dark matter-induced events can present a distinct spectral shape compared to other background contributions at all dark matter masses considered.

\begin{figure}[t]
  \begin{subfigure}{0.49\textwidth}
        \centering
        \includegraphics[scale=0.55]{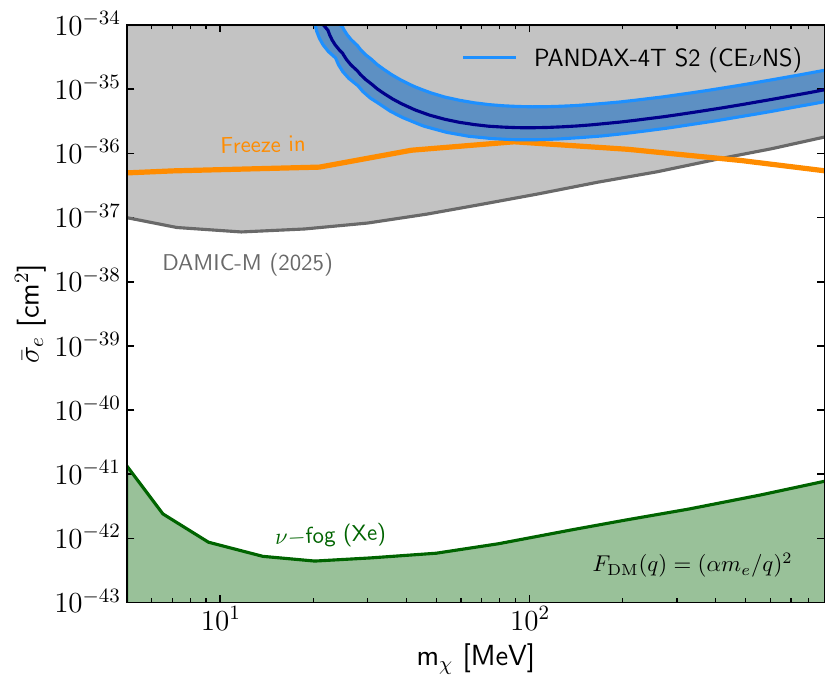}
    \end{subfigure}

     \caption{\justifying Upper limits on the non-relativistic scattering cross section $\sigma_e$, for an ultralight or massless mediator. For comparison, we show complementary bounds from DAMIC-M \cite{DAMIC-M:2025luv}. We further show the relic abundance expectation for the freeze-in mechanism \cite{Hall:2009bx, Essig_2012}, and the neutrino fog from neutrino-electron elastic scatterings \cite{Carew:2023qrj, Wyenberg:2018eyv}.}
    \label{fig:x-section-results-massless}
\end{figure}

From the computation of the dark matter-electron event rate, we turn into deriving upper limits on the dark matter parameters. We define a log-likelihood function summing over all bins $i$ as
\begin{equation}\label{eq:binned_likelihood}
\log \mathcal{L}(\bar{\sigma}_e, m_\chi) = \sum_i \left[ n_i \log(s_i + b_i) - (s_i + b_i) - \log(n_i!) \right],
\end{equation}
where $n_i$ is the number of observed events in the bin $i$, $b_i$ is the number of expected background events (including CE$\nu$NS) in bin $i$, and $s_{i}=s_{i}(\bar{\sigma_e}, m_{\chi})$ is the predicted signal. We then define a test statistic based on the log-likelihood ratio
\begin{equation}
\chi^2(\bar{\sigma}_e, m_\chi) \equiv -2 \log \mathcal{L}(\bar{\sigma}_e, m_\chi).
\end{equation}

and find a $90\%$ C.L upper bound on $\bar{\sigma}_{e}$ from

\begin{equation}
     \chi^{2}(\bar{\sigma}_{e} , m_{\chi}) - \chi^{2}_{\rm min} \leq 2.71\,.
\end{equation}
Under this prescription, we show upper limits on the dark matter-electron scattering cross section from PandaX-4T S2 data, for a heavy mediator in Figure \ref{fig:x-section-results}, and for an ultralight or massless mediator in Figure \ref{fig:x-section-results-massless}. We display the limits from PandaX-4T as shaded blue bands, accounting for different quenching factor models described in the Appendix \ref{app:quenching}. The most stringent limit on the band corresponds to the nominal charge yield model from PandaX-4T, and the least stringent limit corresponds to the Lindhard model. The dashed grey region shows a combination of previous limits from DAMIC-M \cite{DAMIC-M:2025luv}, a previous PandaX analysis \cite{PandaX:2022xqx}, and CRESST-III \cite{CRESST:2019jnq}. Importantly, our novel derived bound from PandaX-4T S2-only data for a CE$\nu$NS search improves over previous PandaX limits \cite{PandaX:2022xqx}. The improvement is mainly driven by a larger (a factor of $\sim 2$) exposure in the CE$\nu$NS search data, and a larger efficiency in the ROI, particularly near threshold. The enhancement in the efficiency increases the number of expected events near threshold significantly, which for a binned-analysis relying on spectral shape differences between the background model and the signal leads to a larger improvement for low dark matter masses than at large dark matter masses. We also find the quenching factor to play an important role, and the best-fit efficiency determination from PandaX-4T leads to more stringent limits than the Lindhard model about an order of magnitude at most, see Appendix \ref{app:quenching} for details.

Further details on the different sub-GeV dark matter models confronted with our bounds and their status in light of recent direct detection results are provided in Section \ref{sec:models}.

\section{Implications for thermal and non-thermal dark matter models}\label{sec:models}

We discuss here the implications of the PandaX-4T S2-channel search for CE$\nu$NS on different thermal and non-thermal sub-GeV dark matter models. All these models share a common feature, meaning an additional $U^{\prime}(1)$ symmetry featuring a vector boson mediator between the Standard Model and the Dark Sector, the so-called ``dark-photon" \cite{holdom1986two}. This has a Lagrangian
\begin{equation}
\begin{aligned}
\mathcal{L}= & -\frac{1}{4} F_{\mu \nu} F^{\mu \nu}-\frac{1}{4} F_{\mu \nu}^{\prime} F^{\prime \mu \nu}-\frac{\epsilon}{2} F_{\mu \nu} F^{\prime \mu \nu} \\
& +\frac{m_{A^{\prime}}^2}{2} A_\mu^{\prime} A^{\prime \mu}+A^{\prime \mu} g_{\chi}J_{\mu}^{\prime}.
\end{aligned}
\label{eq:darkphoton_lagrangian}
\end{equation}
with $F^{\prime \mu \nu}=\partial^\mu A^{\prime \nu}-\partial^\nu A^{\prime \mu}$. What changes primarily among different models is the dark current $J^{\prime}_{\mu}$, together with details on the cosmological histories of $A^{\prime}$ and $\chi$. How the vector boson mass, $m_{A^{\prime}}$ is generated and the relevant kinetic mixing $\epsilon$ value is a model building question. If $m_{A^\prime}$ is generated through dark Higgs mechanism and $\epsilon$ is generated at the loop-level, the available parameter space able to yield the relic abundance of dark matter via s-wave annihilations is largely constrained by bounds on self-interactions \cite{Randall:2008ppe, Markevitch:2003at}, completely ruling out dark photon masses $m_{A^{\prime}} <46$ MeV \cite{Cline:2024wja}.

\subsection{Scalar dark matter}

This model is described by a complex scalar dark matter candidate $\chi$, with dark current in Eq. \ref{eq:darkphoton_lagrangian}.
\begin{equation}\label{eq:scalarcurrent}
J_{\mu}^{\prime}=i \chi^* \partial_\mu \chi
\end{equation}
In this model, the relic abundance can be set via p-wave self-annihilation into electron-positron pairs $\chi\chi^{\star} \rightarrow e^{+}e^{-}$ \cite{Krnjaic:2025noj}. The corresponding parameters yielding the dark matter relic abundance are shown as an orange line in the left panel of Figure \ref{fig:x-section-results}, for two possible ratios of the dark matter to dark-photon mass $\mathcal{R}=m_{A^{\prime}}/m_{\chi}$, and where it is assumed that $A^{\prime}$ decays primarily into dark matter states in the early Universe $A^{\prime} \rightarrow \chi\chi^{\star}$. Similar benchmarks are obtained in the secluded regime
$(2m_{\chi} > m_{A^{\prime}})$), where $A^{\prime}$ decays primarily into electron-positron pairs. Near the resonance regime ($m_{A^{\prime}} \simeq 2m_{\chi}$) self-annihilations are enhanced $ \sigma_{\chi\chi^{\star} \rightarrow e^{+}e^{-}} \sim 1/(s-m_{A^{\prime}}^{2})^{2}$, meaning only $\mathcal{R} \sim 2-3$ this model remains viable in certain regions of parameter space.

In particular, we find that PandaX-4T rules out mediator/dark matter mass ratios larger than $\mathcal{R} \sim 2.8$ for dark matter masses $m_{\chi} \sim 20-200$ MeV. This is an improvement over DAMIC-M and previous Pandax-4T results and in the aforementioned mass range, see the left panel of Figure \ref{fig:x-section-results}. 

\subsection{Asymmetric dark matter}
If the dark matter is a (Dirac) fermion, the current simply reads
\begin{equation}\label{eq:vector_current}
J_{\mu}^{\prime}=\bar{\chi} \gamma^\mu \chi
\end{equation}
and the relic abundance can be set via self-annihilations $\chi\bar{\chi} \rightarrow e^{-}e^{+}$. Bounds from the Cosmic Microwave Background exclude the full sub-GeV dark matter regime in this model \cite{Planck:2018vyg, Slatyer:2015jla}. This only holds for symmetric dark matter. If the dark matter of the Universe is asymmetric, meaning that the particle or anti-particle component was depleted in the early Universe via self-annihilations, then the sub-GeV mass range can still yield the relic abundance for viable dark matter parameters. This model can still lead to dark matter annihilations with the residual component at the time of CMB, which excludes cross sections below the orange line in the right panel of Figure \ref{fig:x-section-results}. For larger values of the non-relativistic cross section $\bar{\sigma}_e$, however, the model remains viable provided the particle-antiparticle asymmetry is sufficiently large.

In light of DAMIC-M results, this model was only viable in two windows in between $m_{\chi}\sim 15-40$ MeV, and $m_{\chi} \sim 150-250$ MeV. We find that the high-mass windows is fully excluded by PandaX-4T, and the small-mass window is only allowed by PandaX-4T in the very narrow range $m_{\chi} \sim 15-20$ MeV.

\subsection{SIMP/ELDER}
Strongly Interacting Massive Particles (SIMP) may couple to the visible sector via kinetic mixing, for instance with an analogous current to the scalar dark matter in Eq. \ref{eq:scalarcurrent}. A key feature of these models is an enlarged dark sector with its distinct decoupling dynamics. For instance, enabling $3 \rightarrow 2$ processes in the dark sector and imposing $\mathcal{O}(1)$ couplings points to an MeV mass scale \cite{Hochberg:2014dra, Hochberg:2014kqa}
\begin{equation}
m_{\chi} \sim \alpha_{\chi}\left(T_{\mathrm{eq}}^2 M_{\mathrm{Pl}}\right)^{1 / 3} \sim 100 \, \mathrm{MeV}
\end{equation}
where $T_{\rm eq}$ is the temperature at matter-radiation equality and $M_{\rm pl}$ is the Planck scale. In this scenario, the dark matter relic density is determined via
processes that change dark matter quantum number. An interesting alternative was proposed in \cite{Kuflik:2015isi}, dubbed ELDER, where the freeze-out happens via elastic scatterings of dark matter particles off Standard Model particles. This mechanism is similar to the SIMP, but here the elastic scattering process decouples first while the self-annihilation process remains active. The range of coupling values that allows to reproduce the relic abundance are in a similar range to SIMPs, corresponding to $\epsilon \sim 10^{-9}-10^{-6}$.

These models yield the correct relic abundance for smaller cross-sections than the scalar and asymmetric dark matter scalar models. Instead of a line, these models can work over a band, which we illustrate in Figure \ref{fig:x-section-results} in purple color, indicating the lower limit from the ELDER mechanism with a dashed black line. We find that PandaX-4T dives well into these models in the mass range from $m_{\chi} \sim 20-300$ MeV, but unable to reach the lower limit from ELDER by about half an order of magnitude.

\subsection{Freeze-in}
If the dark photon mediator is very light or massless, the interaction cross section of dark matter with electrons is momentum squared suppressed $\sigma \sim 1/q^2$. In such a scenario, the dark matter may initially be decoupled from the early Universe plasma, and freeze-in as the Universe expands and colds \cite{Hall:2009bx}. This mechanism sets the relic abundance of sub-GeV dark matter along the orange line in Figure \ref{fig:x-section-results-massless} \cite{Hall:2009bx,Essig_2012}. The dark current can be analogous to Eq. \ref{eq:scalarcurrent} and Eq. \ref{eq:vector_current}. 

This model is robustly tested by DAMIC-M. We find that PandaX-4T is also able to probe the model in the mass range $m_{\chi} \sim 70-130$ MeV, further lying within a factor of $\sim 2$ for a wider range of masses. This model remains only viable in the very low mass range $m_{\chi} \lesssim 3$ MeV, or in the high mass range $m_{\chi} \gtrsim 400$ MeV.

\subsection{$L_{\mu}-L_{\tau}$ symmetry}
In this model the additional U(1)$^{\prime}$ symmetry is gauged in the differences of muon and tau lepton numbers $L_{\mu}-L_{\tau}$ \cite{He:1990pn,He:1991qd,Foot:1990mn}. The dark current is analogous to Eq. \ref{eq:vector_current}, but importantly, the muon and tau lepton doublets also possess a current

\begin{equation}
\mathcal{L} \supset g_{\mu \tau}\left(\bar{\mu} \gamma_\mu \mu-\bar{\tau} \gamma_\mu \tau+\bar{\nu}_\mu \gamma_\mu P_L \nu_\mu-\bar{\nu}_\tau \gamma_\mu P_L \nu_\tau\right) {A^{\prime}}^{\mu},
\end{equation}
and the size of the kinetic mixing $\epsilon$ can be predicted at the loop-level as
\begin{equation}\label{eq:loop_suppression}
\epsilon=\frac{e g_{\mu \tau}}{12 \pi^2} \log \frac{m_\tau^2}{m_\mu^2} \simeq g_{\mu \tau} / 70 .
\end{equation}
The dark gauge coupling is $g_{\chi}=g_{\mu \tau}Q_\chi$, and $Q_\chi$ is the charge of the dark matter under the $U(1)_{L_\mu-L_\tau}$ symmetry.

\begin{figure}[t]
  \begin{subfigure}{0.49\textwidth}
        \centering
        \includegraphics[scale=0.53]{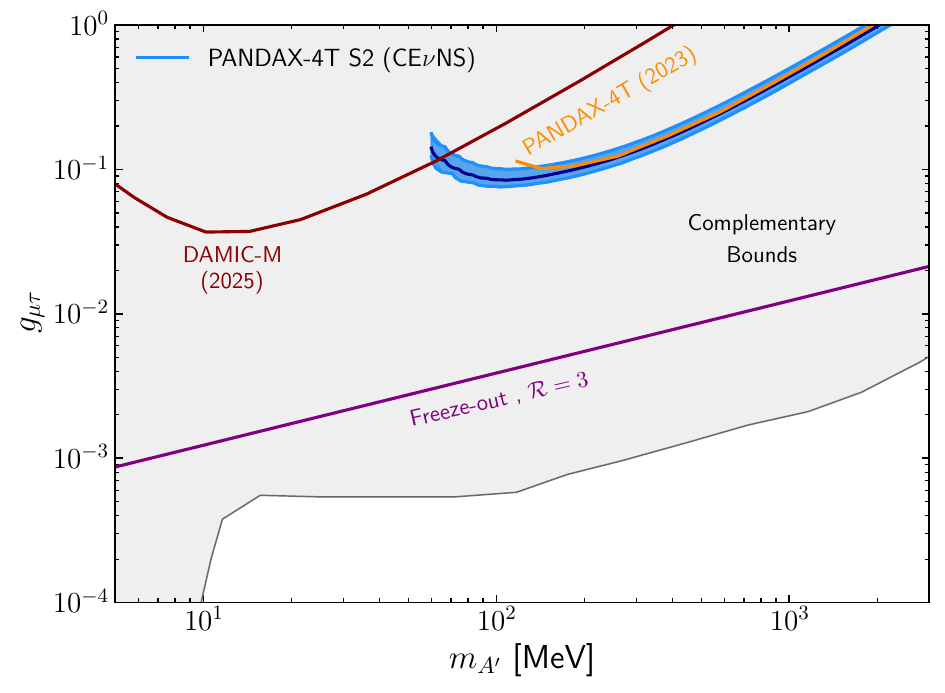}
    \end{subfigure}

     \caption{\justifying Upper limit on $g_{\mu \tau}$ coupling as function of the mediator mass $m_{A^{\prime}}$ , for $\mathcal{R} =3$ and considering $g_{\chi} = g_{\mu \tau}$. We use $\epsilon \simeq g_{\mu \tau} / 70$, see Eq. \ref{eq:loop_suppression}. The purple line corresponds to the combination of values able to reproduce the observed relic abundance of dark matter, and the grey line corresponds to a combination of accelerators and cosmological bounds, taken from \cite{bernal2025enablingsubgevdarkmatter, PhysRevD.105.016014, Kahn_2018, Biswas:2016yan}.}
     \label{fig:limit-lmultau}
\end{figure}

For sub-GeV dark matter, this model remains widely untested by direct detection \cite{Foldenauer:2018zrz,Holst:2021lzm,Hapitas:2021ilr,figueroa2024directdetectionlightdark, Bernal:2025szh}. Bounds from accelerators are generically stronger than those from direct detection in the regime $m_{A^{\prime}} >2m_{\chi}$. The recent results from PandaX-4T bring them closer to accelerator bounds, but still unable to test thermal freeze-out. We show upper limits on the gauge coupling $g_{\mu \tau}=g_{\chi}$ as a function of the mediator mass $m_{A^{\prime}}$ from DAMIC-M and PANDAX-4T in Figure \ref{fig:limit-lmultau}, confronted with complementary bounds from accelerators and measurements of $N_{\rm eff}$ at BBN in this model \cite{Foldenauer:2018zrz}, and the region of parameter space able to yield the relic abundance via thermal freeze-out, for $\mathcal{R}=3$ \cite{Foldenauer:2018zrz, Hapitas:2021ilr}. Importantly, the recent bound on the anomalous moment of the muon $g-2$ seems to exclude to the freeze-out benchmark in this model at the MeV scale for $\mathcal{R}=3$ \cite{Muong-2:2025xyk,Bernal:2025szh}. However, values closer to the resonance regime are still allowed. It can be noticed that current direct detection bounds are still weaker than those from accelerators in the parameter $g_{\mu \tau}$ by a factor of at least $\gtrsim 5$, depending on the mass considered. This is mainly due to the loop suppression in $\epsilon$ mentioned in Eq. \ref{eq:loop_suppression}. It should be highlighted that the Planck constraint on $N_{\rm eff}$ in this model restricts gauge couplings below $g_{\mu \tau} \lesssim 10^{-4}$ for $m_{A^{\prime}} \lesssim 20$ MeV, thus being very restrictive.

The improvement from DAMIC-M and PandaX-4T turns out more relevant in the secluded annihilation regime, $m_{A^{\prime}}<2m_{\chi}$, where the relic abundance is set by annihilations $\bar{\chi}\chi \rightarrow A^{\prime}A^{\prime}$. The value of $g_{\mu\tau}$ needs to lie below $g_{\mu\tau} \sim 5 \times 10^{-4}$, otherwise the model could be in conflict with the anomalous magnetic moment of the muon $g-2$ \cite{Muong-2:2025xyk,Li:2025myw}. In this case, PandaX-4T places limits that lie less than a factor of $\sim 2$ away from the relic abundance targets, and DAMIC-M lies away by a factor of $\sim 10$ \cite{figueroa2024directdetectionlightdark}.

\subsection{Majorana}
In this model the dark matter is a Majorana fermion with current
\begin{equation}
J^{\prime}_{\mu}=\frac{1}{2} \bar{\chi} \gamma_\mu \gamma_5 \chi.
\end{equation}
This model is poorly tested by direct detection due to the suppression of the scattering cross section with relative velocity as $\sigma \sim v^2$, since the typical dark matter velocity in the galactic halo is $v \sim 10^{-3}$c. On the other hand this model is better probed by accelerators and relativistic dark matter probes, as cosmic ray boosted dark matter or cosmic ray cooling in Active Galactic Nuclei. We comment on these alternative phenomenological probes in Section \ref{sec:complementary}. The strength of direct detection and complementary probes, and their comparison to thermal freeze-out, is shown in Fig. \ref{fig:x-section-results-majorana}.

\subsection{Inelastic dark matter}
This model features a mass splitting among two dark matter states $\chi_1$ and $\chi_2$, and off-diagonal interactions at the tree-level, while diagonal interactions are loop-suppressed. The current is
\begin{equation}
J_{\mu}^{\prime}=i \bar{\chi}_1 \gamma_\mu \chi_2.
\end{equation}

This model is also generically better constrained by accelerators than direct detection for mass splittings larger than $m_{\chi_2}-m_{\chi_1} \gtrsim $ 500 keV, but the parameter space yielding the thermal relic abundance remains untested at the MeV scale in some portions of parameter space \cite{Berlin:2018bsc,CarrilloGonzalez:2021lxm,Garcia:2024uwf, DallaValleGarcia:2024zva}. Bounds from cosmic ray cooling in Active Galactic Nuclei and cosmic ray boosted dark matter may test this scenario better than colliders for sub-GeV dark matter, see Section \ref{sec:complementary} for further details.

\section{Complementary constraints from direct detection, astrophysics, cosmology and beam dumps}\label{sec:complementary}

In Figures \ref{fig:x-section-results} and \ref{fig:x-section-results-massless}, we have only presented our limits from PandaX-4T S2-channel search for CE$\nu$NS with other direct detection bounds on dark matter under the SHM model, to allow for a fair comparison. However, there are several other complementary bounds on sub-GeV dark matter that can be relevant in certain regions of parameter space. We discuss some of them qualitatively in the following.
\subsection{Nuclear recoils and the Migdal effect}
The Lagrangian from Eq. \ref{eq:darkphoton_lagrangian} also leads to dark matter-proton interactions through kinetic mixing. The spin-independent non-relativistic dark-matter proton scattering cross section in the heavy mediator limit $\sigma_p$ can be rescaled to the non-relativistic dark matter-electron scattering cross section $\bar{\sigma_e}$ as
\begin{equation}
\sigma_{p} = (\mu^{2}_{\chi p} / \mu^{2}_{\chi e})\bar{\sigma}_{e}
\end{equation}
In the sub-GeV dark matter mass regime, the leading experiments probing dark matter-induced nuclear recoils are CRESST-III and SuperCDMS \cite{CRESST:2019jnq, SuperCDMS:2014cds}. We find that CRESST-III nuclear recoil data becomes important in testing sub-GeV dark matter models featuring a heavy mediator. For intance, for the asymmetric dark matter case, nuclear recoil data excludes the thermal region better than PandaX-4T for $m_{\chi} \gtrsim 250$ MeV.

In addition to nuclear recoil searches, it has been discussed in recent years that dark matter-induced nuclear recoils would also yield an additional ionization signal arising from the Migdal effect \cite{Ibe:2017yqa, Dolan:2017xbu, Baxter:2019pnz}. The ionization signal from the Migdal effect is expected to occur at larger electron-equivalent energy than the original nuclear recoil energy in electron equivalent energy. This may help to extend the sensitivity to light dark matter and to test sub-GeV dark matter models with large-scale exposure experiments \cite{XENON:2019zpr, DarkSide:2022dhx}. The Migdal effect signal from solar neutrino CE$\nu$NS also constitutes an additional neutrino ``fog" for sub-GeV dark matter searches \cite{Herrera:2023xun, Maity:2024hzb}. This becomes particularly relevant in the high mass range $m_{\chi} \sim 0.1-1$ GeV.

\begin{figure}[t]
  \begin{subfigure}{0.49\textwidth}
        \centering
        \includegraphics[scale=0.7]{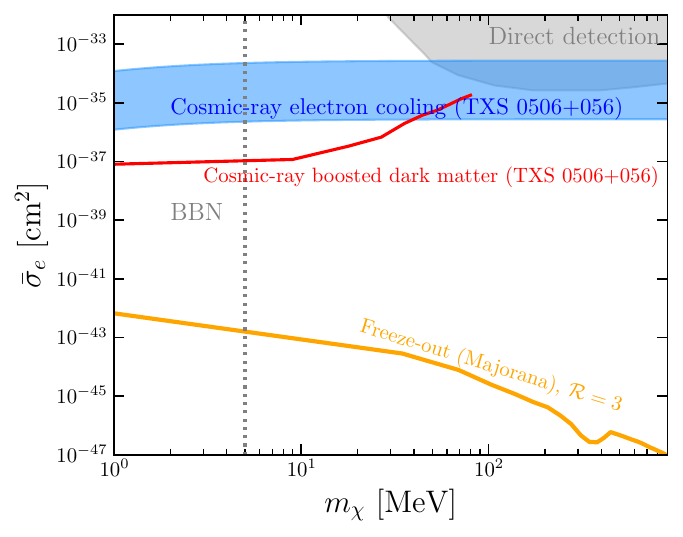}
    \end{subfigure}

     \caption{\justifying Bounds on Majorana sub-GeV dark matter from direct detection of galactic halo dark matter at PANDAX-4T (grey, this work), direct detection of cosmic ray boosted dark matter at Super-Kamiokande (red) \cite{Granelli:2022ysi}, cosmic ray cooling from Active Galactic Nuclei (blue) \cite{Herrera:2023nww}, and BBN (dotted grey line) \cite{Krnjaic:2025noj}. We confront the results with the thermal freeze-out benchmark (orange) \cite{Krnjaic:2025noj}.}
    \label{fig:x-section-results-majorana}
\end{figure}
\subsection{Non-standard dark matter velocity distributions}

The Standard Halo Model is believed to be a good approximation of the galactically-bounded dark matter velocity distribution. This is supported by a combination of N-body simulations and observational methods relying on stellar tracers, which however find some sizable deviations from the SHM, \textit{e.g} \cite{Vogelsberger:2008qb,Schaye:2014tpa,Bozorgnia:2016ogo,Evans:2018bqy,Necib:2018iwb,OHare:2018trr,Folsom:2025lly}. These deviations are small and do not change the limits for dark matter-electron scattering searches by more than one order of magnitude near the kinematical threshold, and a factor of $\sim 2$ at higher dark matter masses \cite{Bozorgnia:2019mjk,Radick:2020qip,Maity:2022enp,Hryczuk:2020trm,Herrera:2024zrk}.

The effect of unbounded dark matter particles could be more significant in dark matter-electron scattering searches \cite{Herrera_2021, Herrera:2023fpq, Smith-Orlik:2023kyl, Kakharov:2025myy, DeBrae:2025umb, Santos-Santos:2023ubx, Reynoso-Cordova:2024xqz}. These unbounded particles could arise, for instance, from the the Local Group, or from the Large Magellanic cloud. These particles, although with density smaller than those bounded to the galactic halo, have velocities larger than the escape velocity of the Milky Way, thus enhancing the sensitivity near the threshold of dark matter searches.

The enhancement from the Local Group component in dark matter-electron scattering limits near threshold can be as large as $\sim 2$ orders of magnitude, further extending the sensitivity to smaller dark matter masses \cite{Herrera_2021}. The impact for the Large Magellanic Cloud is expected to be smaller, of at most about an order of magnitude, but it is robustly supported by simulations \cite{Smith-Orlik:2023kyl}. This could be particularly relevant to close the yet allowed window at low masses in the asymmetric dark matter model presented in Fig. \ref{fig:x-section-results}, and the small mass window allowed for the freeze-in model presented in Fig. \ref{fig:x-section-results-massless}.

\subsection{Cosmic-ray boosted dark matter}
Cosmic rays may scatter off sub-GeV dark matter in the galactic halo or in extragalactic sources, inducing a flux of boosted dark matter particles on Earth
\cite{Bringmann:2018cvk}. For sub-GeV dark matter-electron interactions, the limits from the galactic cosmic ray electron component are weak, in the range $\bar{\sigma}_{e} \sim 10^{-30}-10^{-28}$cm$^2$ for $m_{\chi} \sim 1-1000$ MeV for a heavy $m_{A^{\prime}}$ \cite{Herbermann:2024kcy}. When considering a boosted component from the blazar TXS 0506+056, the sensitivity is enhanced by $\sim 7$ orders of magnitude compared to the galactic case \cite{Granelli:2022ysi}. These limits are comparable in the low-mass range to those from DAMIC-M for the galactically bound dark matter, but rely on assumptions on the cosmic ray electron spectra at the source, which presents large uncertainties.

The boosted dark matter component can be semi-relativistic or relativistic, which becomes relevant for probing Majorana dark matter models suppressed by the relative velocity squared $v^2$, see Fig. \ref{fig:x-section-results-majorana}, and inelastic dark matter models with mass splittings larger than $\sim 500$ keV \cite{Bell:2021xff,CRBDM}.

\subsection{Solar-reflected dark matter}
Halo dark matter may scatter in the hot plasma of the Sun, gaining energy in the process. This process occurs at a sufficient large rate to yield a sizable "solar reflected" dark matter flux on Earth, which enhances the sensitivity to sub-GeV dark matter in the lowest mass end \cite{Emken:2017hnp,An:2021qdl, Emken:2024nox}. A recent dedicated analysis from the PandaX-4T collaboration \cite{PandaX:2024syk} excludes cross sections as low as $\bar{\sigma}_{e} \sim 5 \times 10^{-39}$cm$^2$ in the range $m_{\chi} \sim 0.01-10$ MeV, for a heavy mediator. The majority of this parameter space is already excluded by measurements of $N_{\rm eff}$ at BBN and CMB, and the allowed region at larger masses is better probed by DAMIC-M by about half an order of magnitude.

\subsection{Cosmic-ray cooling in Active Galactic Nuclei and Starburst Galaxies}
Multi-messenger observations from some Active Galactic Nuclei (AGN) indicate that cosmic rays are efficiently cooled, leading to neutrino and gamma-ray emission. The ambient dark matter in these sources can scatter off cosmic rays, altering the neutrino and gamma-ray emissions \cite{Herrera:2023nww, Mishra:2025juk, DeMarchi:2024riu, DeMarchi:2025xag}. Data from NGC 1068 and TXS 0506+056 have been used to place stringent limits on sub-GeV dark matter, with similar strength to those from solar reflection for $m_{\chi} \lesssim 50$ MeV. Similar bounds have been derived from Starburst Galaxies \cite{Ambrosone:2022mvk}, yet probing somewhat larger cross sections.

Cosmic ray cooling in AGN is particularly relevant to probe Majorana dark matter models. In Figure \ref{fig:x-section-results-majorana}, we show a band accounting for a range of possible limits that are derived from TXS 0506+056 when accounting for astrophysical uncertainties on the cosmic ray proton luminosity at the source, the precision of high-energy neutrino and gamma-ray measurements, and uncertainties on the dark matter density profile at the source \cite{Herrera:2023nww}. Cosmic-ray cooling is also useful to probe for inelastic dark matter models, where the ensuing limits can be comparable or stronger to those from accelerators \cite{Gustafson:2024aom} for mass splittings $m_{\chi_2}-m_{\chi_1}\sim 0.1-1 m_{\chi_2}$.

\subsection{CMB and BBN}

If in equilibrium with the SM plasma at early times, dark matter annihilations or scatterings may increase the effective number of relativistic species $N_{\rm eff}$ at BBN and recombination, which is constrained by observations of primordial abundances and CMB anisotropies \cite{Sabti:2019mhn,Giovanetti:2021izc}. These limits vary with the dark matter candidate and mediator mass, due to the different degrees of freedom in the model and cross sections. They typically exclude $m_{\chi} \lesssim 5-15$ MeV down to very small cross sections, as indicated in the panels of Fig. \ref{fig:x-section-results} as a vertical dotted grey line.

The CMB energy injection bounds from dark matter annihilations also exclude symmetric dark matter candidates with sizeable s-wave annihilation cross sections below a few GeV \cite{Madhavacheril:2013cna,Slatyer:2015jla}, and even asymmetric dark matter~\cite{Graesser:2011wi,Lin:2011gj}. In particular, defining the fractional asymmetry as $r_{\infty}
\equiv \Omega_{\bar{\chi}}/\Omega_{\chi}$ the constraint is~\cite{Lin:2011gj}

\begin{equation}
\langle\sigma v\rangle_{\mathrm{CMB}}\lesssim \frac{2.42 \times 10^{-27} \mathrm{~cm}^3 / \mathrm{s}}{2 f}\left(\frac{m_{\chi}}{1 \, \mathrm{GeV}}\right)\left(\frac{1}{r_{\infty}}\right)
\end{equation}
where $f$ gives the efficiency of energy deposition at redshift $z$ and thus depends on the spectrum photons, neutrinos and $e^{ \pm}$resulting from light dark matter annihilation, and takes values $f \sim 0.2-1$. This bound, combined with the relic abundance constraint on asymmetric DM~\cite{Graesser:2011wi}, excludes the region below the orange line in the right panel of Fig. \ref{fig:x-section-results}.

\subsection{Dark matter self-annihilations in galaxies}
In addition to constraints from energy injection from sub-GeV dark matter annihilation in the CMB, there are constraints from sub-GeV dark matter annihilations in the galactic halo \cite{Essig:2013goa,Boudaud:2016mos,Cirelli:2020bpc, Cirelli:2023tnx}. Sub-GeV dark matter annihilations can contribute to the diffuse X-ray and gamma-ray observations. These bounds are weaker than those from the CMB for s-wave annihilation channel for $m_{\chi} \lesssim 200$ MeV, but they can be stronger for heavier dark matter.

\subsection{Accelerators}
Sub-GeV dark matter can be produced at both proton and electron beam dump experiments, then scattering at the nearby detector. Sub-GeV dark matter can also be produced at $e^{+}e^{-}$ colliders yielding missing energy signatures. The strongest constraints arise from a combination of NA64 and BaBar \cite{NA64:2024klw,Essig:2013vha}. These constraints are most relevant in those models where the scattering cross section of dark matter with electrons is kinematically-suppressed or loop-suppressed, such as Majorana or Inelastic dark matter scenarios, or concrete models such as the $L_{\mu}-L_{\tau}$. However, they are yet not able to probe large portions of the parameter space favored for thermal production. Future experiments like Belle-II and LDMX may be able to exclude larger regions of the parameter space \cite{LDMX:2018cma}.

\subsection{Supernovae}
Sub-GeV dark matter can be produced in the dense core of supernovae, escaping and carrying away energy. This alters the properties of the neutrino burst, for instance in SN1987A \cite{Chang:2018rso}. The ensuing constraints on sub-GeV dark matter are strong for feeble couplings, excluding kinetic mixings in the range of $\bar{\sigma}_e \sim 10^{-47}-10^{-40}$cm$^2$. This window corresponds to smaller cross sections than most of the well-motivated models presented in Section \ref{sec:models}, however. In the case of massless mediators, the limits are not so stringent, in fact they are weaker than those from DAMIC-M and PandaX-4T, except for $m_{\chi} \lesssim 3$ MeV.

\section{Conclusions}\label{sec:discussion}
The DAMIC-M collaboration recently reported strong results on sub-GeV dark matter, robustly testing large portions of the parameter space of thermal and non-thermal models. We have derived novel bounds from the recent PandaX-4T S2-only ionization search for CE$\nu$NS, finding that these are world-leading for heavy mediators in the range $m_{\chi} \sim 20-200$ MeV, but weaker than DAMIC-M for light or massless mediators. PandaX-4T puts significant pressure on several sub-GeV dark matter models, nearly fully excluding the whole parameter space of asymmetric Dirac dark matter fermions, and fully excluding scalar dark matter with $m_{A^{\prime}} \gtrsim 3m_{\chi}$. Besides testing these simple benchmark scenarios, the recent data from DAMIC-M and PandaX-4T also dives well into other models such as SIMPs/ELDERs, and approach benchmarks for freeze-in, and concrete gauged symmetries such as the $U(1)_{L_{\mu}-L_{\tau}}$.

We have assessed the uncertainty in the PandaX-4T models arising from different choices of the charge-yield and quenching factors. The best-fit model from PandaX-4T yield limits that can be more stringent than those obtained for the Lindhard model by a factor of $\sim 2$.

In addition to deriving novel bounds from PandaX-4T S2-only ionization data for CE$\nu$NS, we have contextualized the landscape of sub-GeV dark matter with direct detection probes, astrophysics, cosmology and other laboratory probes. We paid particular attention to detection methods that directly probe scatterings between the dark matter and electrons, either in the Lab or in astrophysical set-ups. These complementary searches can help DAMIC-M and PandaX-4T in probing some models, for instance Majorana dark matter, inelastic dark matter, or dark matter charged under the $L_{\mu}-L_{\tau}$-symmetry, where the direct detection of galactic halo dark matter is kinematically or loop-suppressed.

There is room to improve the results from DAMIC-M and PandaX-4T. Future CCD-based detectors such as OSCURA \cite{aguilararevalo2022oscuraexperiment} may reach lower thresholds and larger exposures than DAMIC-M, and large-scale liquid xenon detectors like XLZD \cite{XLZD:2024nsu} will also improve over PandaX-4T exposures in the incoming years. Novel detector technologies and large scale neutrino detectors may potentially also help to extend the sensitivity to sub-GeV dark matter, \textit{e.g} \cite{Hochberg:2015pha,Schutz:2016tid,Kurinsky:2019pgb,Kahn:2021ttr,Cogswell:2021qlq,Zema:2024epe,Leane:2025efj}. Narrowing down uncertainties on the charge yields/quenching factors and on the tail of the dark matter velocity distribution are also crucial to robustly test some sub-GeV dark matter models. It is plausible to close most of the yet allowed windows in the parameter space of sub-GeV dark matter within a few years, potentially only leaving space for Majorana and Inelastic dark matter models, which will likely require the use of astrophysics and accelerators.

\section{Acknowledgments}
G.H is grateful to Robert McGehee for useful discussions. AC is supported by the National Natural Science Foundation of China (NSFC) through the grant No.12090064. The work of G.H and I.M.S is supported by the U.S. Department of Energy under the award number DE-SC0020262.

\section{Note added}

While this work was being completed, the PandaX-4T collaboration released an analysis of sub-GeV dark matter from their recent ionization S2-only data \cite{Zhang:2025ajc}. Our analysis is complementary to theirs. Our limits for heavy mediators agree well with theirs, but our limits for light mediators are less stringent by a factor of $\sim 2$. The collaboration explicitly mentions that their limits for light mediators rely on a different treatment on the expected signal than the prescription we described in our Appendix \ref{app:quenching}. Further details would be needed to be able to reproduce the collaboration results. We provide a detailed description on the method used to set upper limits, discuss and compute the sensitivity near the kinematical threshold, and assess the impact of quenching factors. We further contextualize the results in the landscape of the sub-GeV dark matter models and complementary probes in more depth.
\newpage

\bibliography{nu.bib}
\clearpage
\appendix
\onecolumngrid
\section{Details on the charge yield, quenching factors and efficiency functions at PandaX-4T}\label{app:quenching}
Dark matter-electron interactions induce events in electron-equivalent energy keV$_{ee}$. However, the experimental observable is the number of observed electrons $n_{e^{-}}$. To convert ionization energies $E_e$ to number of electrons $n_e^{-}$, we follow Eq. \ref{eq:conversion_ne} in the main text. The number of electrons $n_{e^{-}}$ and (unobservable) scintillation photons read \cite{Essig_2017}

\begin{align}
& n_\gamma=N_{\mathrm{ex}}+f_R N_i, \\
& n_{e^{-}}=\left(1-f_R\right) N_i .
\end{align}
$N_i$ refers to the number of ions, and $N_{\rm ex}$ to the number of excited atoms. In the modified Thomas-Imel recombination model, $f_R=0$ and $n_{e^{-}}=N_i$ and $n_{\gamma}=N_{\rm ex}$, and the fraction of initial quanta observed as electrons is $f_e=\left(1-f_R\right) /\left(1+N_{\mathrm{ex}} / N_i\right) \simeq 0.83$ \cite{Aprile:2007qd,Sorensen:2011bd}. We further take $W=13.8$ eV. For each electron ionization energy $E_e$, the number of additional quanta created is
\begin{equation}
n^{(1)}=\mathrm{Floor}(E_{e}/W)
\end{equation}
We further assume that the photons fron the de-excitation of the following shells $(5 s, 4 d, 4 p, 4 s)$ can photoionize to create an additional quanta $n^{(2)}=\left(n_{5 s}, n_{4 d}, n_{4 p}, n_{4 s}\right)=(0,4,6-10,3-15)$. The total number of electrons is then given by
\begin{equation}
n_{e^{-}} = n_{e^{-}}' + n_{e^{-}}'',
\end{equation}
where $n_{e^{-}}^{'}$ represents the number of primary electrons and $n_{e^{-}}^{''}$ the number of secondary electrons.
The primary electron count is a Bernoulli-distributed variable
\begin{equation}
n_{e^{-}}' =
\begin{cases}
1 & \mathrm{with \, probability }\, f_R, \\
0 & \mathrm{with \,  probability }\, 1 - f_R.
\end{cases}
\end{equation}
The secondary electrons $n_{e^{-}}^{''}$ follow a binomial distribution with $n^{(1)} + n^{(2)}$ trials and a success probability $f_e$
\begin{equation}
P(n_{e^{-}}'' = k) = \binom{n^{(1)} + n^{(2)}}{k} f_e^k (1 - f_e)^{n^{(1)} + n^{(2)} - k}, \quad \mathrm{for } \, k = 0, 1, \dots, n^{(1)} + n^{(2)}.
\end{equation}

For the right panel of \autoref{fig:counts-PandaX-DMe} we have performed an interpolation of the binomial distribution, so that it can be evaluated at half-integers, and match the binning reported in \cite{pandaxcollaboration2024indicationsolar8bneutrino}.

In addition to converting $E_e$ to $n_{e^{-}}$, we need to account for the experimental efficiencies of the experiment in their S2-only ROI, usually reported in nuclear recoil energy $E_{R}$. The first component of the total efficiency arises from the selection efficiency, which is flat in the ROI of the experiment and amounts for about $40\%$ of the events. We account for the reported selection efficiency by PandaX-4T in our analysis. The second factor arises from the ROI efficiency, which is energy dependent. This efficiency needs to be converted into electron equivalent ionization energy $E_e$, or number of electrons $n_{e^{-}}$. The translation depends on the charge yield or quenching factor of the detector, which is uncertain at low energies. We first use the Lindhard model to convert nuclear recoil energy into electron recoil energy. The quenching factor reads \cite{osti_4701226}
\begin{equation}
Q_L\left(E_R, k\right)=\frac{k g\left(E_R\right)}{1+k g\left(E_R\right)}
\end{equation}
where 
\begin{equation}
g\left(E_R\right)=  3\left(11.5 Z^{-7 / 3} E_R\right)^{0.15}+0.7\left(11.5 Z^{-7 / 3} E_R\right)^{0.6}
 +11.5 Z^{-7 / 3} E_R.
\end{equation}
where $Z=54$ for Xenon. We also extract the nominal and best fit charge yields at PandaX-4T, reported in Fig. 4 of \cite{pandaxcollaboration2024indicationsolar8bneutrino} . We then use \cite{De_Romeri_2025}:

\begin{equation}\label{ne_nr_translation}
    n_{e^{-}} = E_{R} Q_{y}(E_{R})
\end{equation}

\begin{figure}[h]
    \centering
    \includegraphics[scale=0.49]{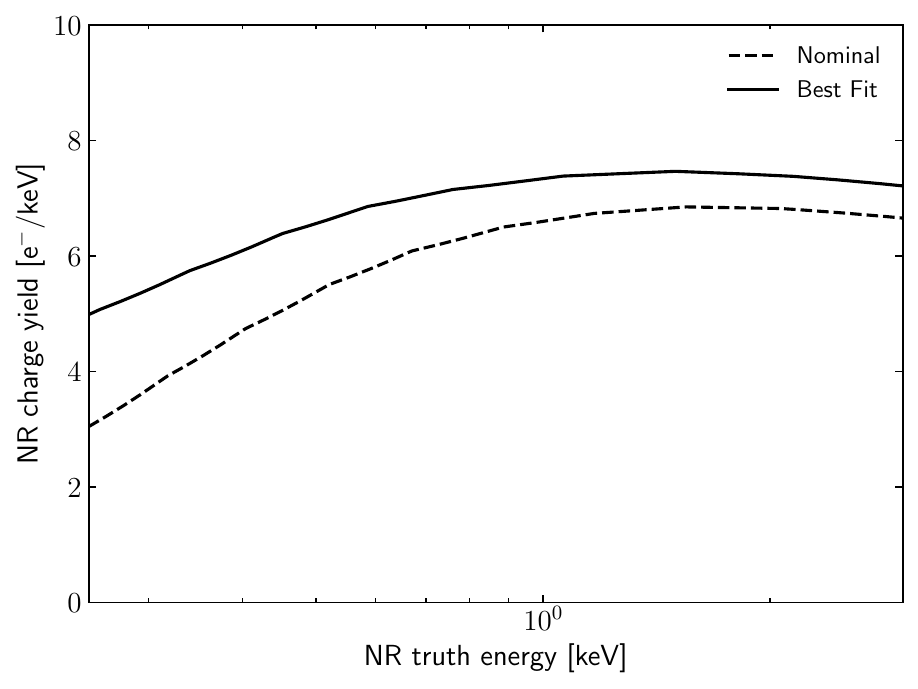}
    \includegraphics[scale=0.49]{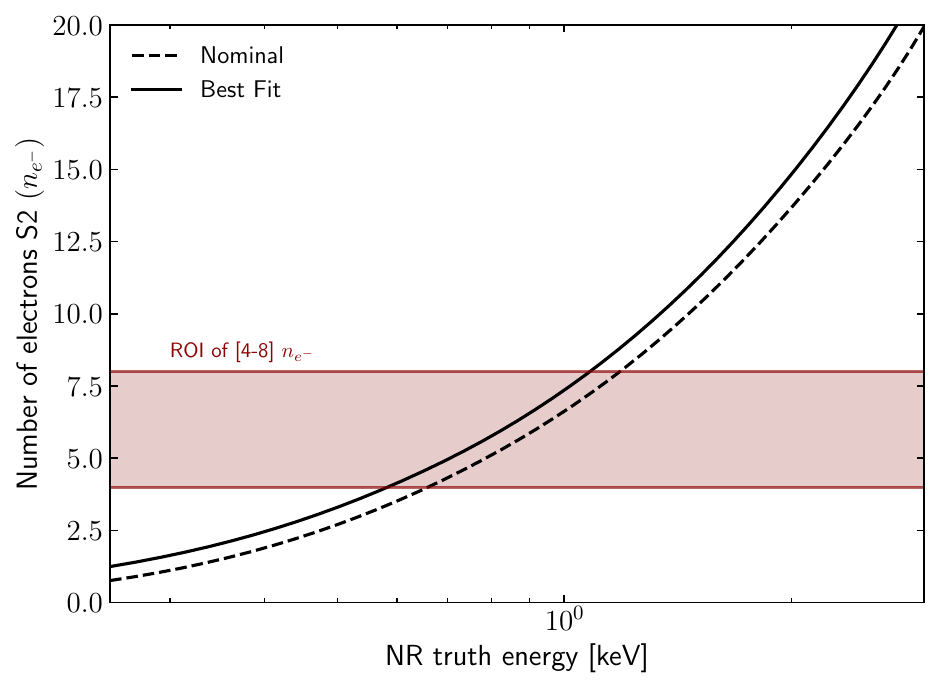}
    \caption{\justifying \textit{Left panel:} Best-fit and Nominal charge yields at PandaX-4T \cite{pandaxcollaboration2024indicationsolar8bneutrino}. \textit{Right panel:} Number of electrons vs nuclear energy following the PandaX-4T charge yields.}
    \label{fig:charge-yield}
\end{figure}
In Figure \ref{fig:charge-yield}, we display the difference among the nominal and best-fit charge yields at PandaX-4T, and the associated impact on the conversion from nuclear recoil energies to number of electrons. Within the ROI, the impact is very small.

\begin{figure}[h]
    \centering
    \includegraphics[scale=0.49]{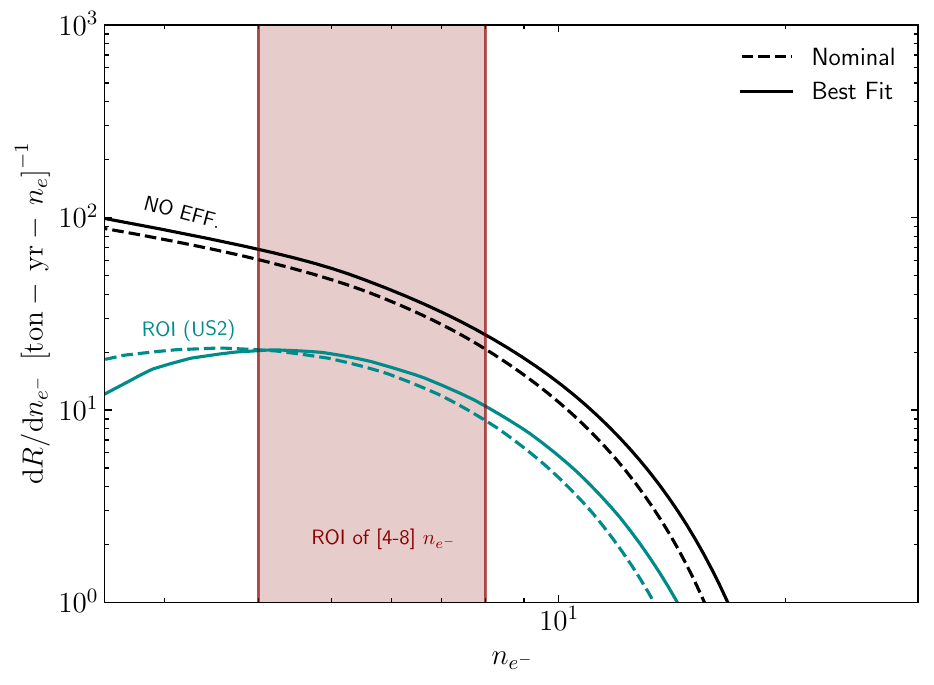}
    \includegraphics[scale=0.49]{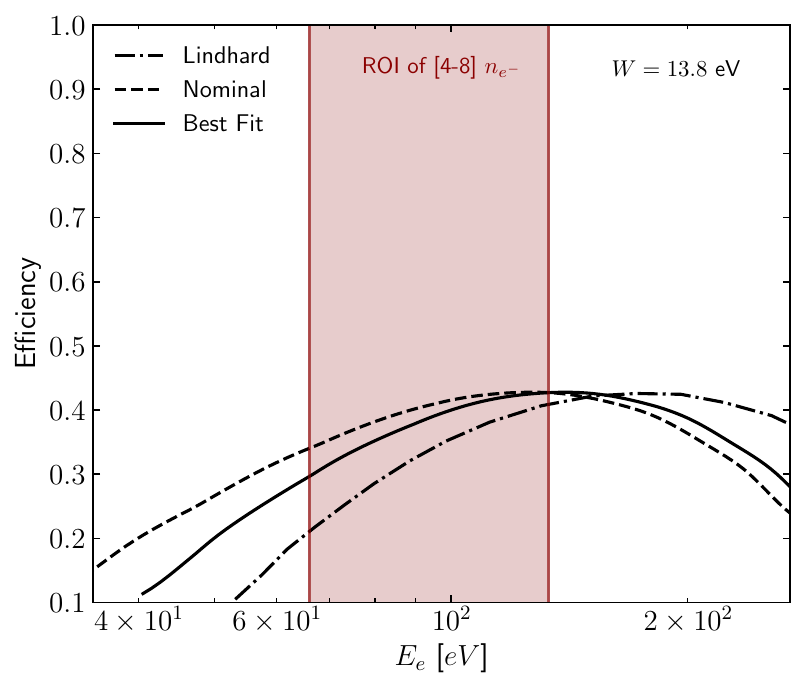}
    \caption{\justifying \textit{Left panel:} $dR/dn_{e^{-}}$ as a function of $n_{e^{-}}$. \textit{ Right panel:} ROI Efficiency function $\mathcal{A}(E_{e})$ for three different quenching models. In addition to the ROI efficiency in the S2 channel, we further consider the selection efficiency, amounting to $40\%$}
    \label{fig:efficiencies}
\end{figure}

In the left panel of Figure \ref{fig:efficiencies}, we show the difference in the differential event rate when including the efficiencies of PandaX-4T versus the rate without including the efficiencies, and the difference among the nominal and best-fit cases. Within the ROI, the impact of the efficiencies is large, of about a factor of $\sim 5$. In the right panel of Figure \ref{fig:efficiencies}, we show the efficiency as a function of electron ionization energy for different conversion models used from nuclear recoil energy to electron equivalent energy: Lindhard (dashed-dotted line), Nominal (dashed line) and best fit (solid). The differences within the ROI are significant at low energy, which are amplified non-linearly in the ionization rate integral, and subsequently in the low-energy bins of the likelihood analysis in Eq. \ref{eq:binned_likelihood}. The uncertainty arising from the different choices is captured by the blue band in the upper limits from Figures \ref{fig:x-section-results} and \ref{fig:x-section-results-massless}. The uncertainty is larger at low masses than at high dark matter masses, as the uncertainties from the efficiencies are more prominent there.

\section{Reproducing the expected CE$\nu$NS rate at PandaX-4T}\label{app:sevens}

Since we make use of the CE$\nu$NS search data at PandaX-4T, we attempt to reproduce the expected event rate in the S2-channel for consistency. A similar task was recently performed in \cite{blancomas2024clarityneutrinofogconstraining, DeRomeri:2024iaw}, finding overall good agreement with PandaX-4T expectations within the ROI, see in table~\ref{tab:recoilexcess}. We compute the CE$\nu$NS scattering rate as

\begin{equation}\label{rate_er_eff}
    \frac{dR}{dE_{R}} = \mathcal{A}(E_{R}) \int dE_{\nu} \frac{d \phi}{dE_{\nu}} \frac{d \sigma_{\nu N}}{dE_{R}}
\end{equation}

where $\frac{d \sigma_{\nu N}}{dE_{R}}$ 
refers to the SM $\nu-N$ cross section, and $\mathcal{A}(E_{R})$ is the experiment-dependent efficiency, which we discussed in the previous Section of the Appendix. We make use of the best-fit charge yield from \cite{pandaxcollaboration2024indicationsolar8bneutrino}. We take the default flux of $^{8}$B solar neutrinos, $d \phi/dE_{\nu}$ with normalization $5.46\times10^{6}\,\mathrm{cm}^{-2}\,\mathrm{s}^{-1}$ from S$\nu$DD \cite{Amaral:2023tbs}.

The data in \cite{pandaxcollaboration2024indicationsolar8bneutrino} is presented in bins of S2 number of electrons ($n_{e^{-}}$). As discussed, the translation between nuclear recoil energy and $n_{e^{-}}$ is carried out through the charge yield $Q_{y}(E_{R})$ as \cite{DeRomeri:2024iaw}

\begin{equation}\label{ne_nr_translation}
    n_{e^{-}} = E_{R} Q_{y}(E_{R}).
\end{equation}

\begin{figure}
    \centering
    \includegraphics[scale=0.43]{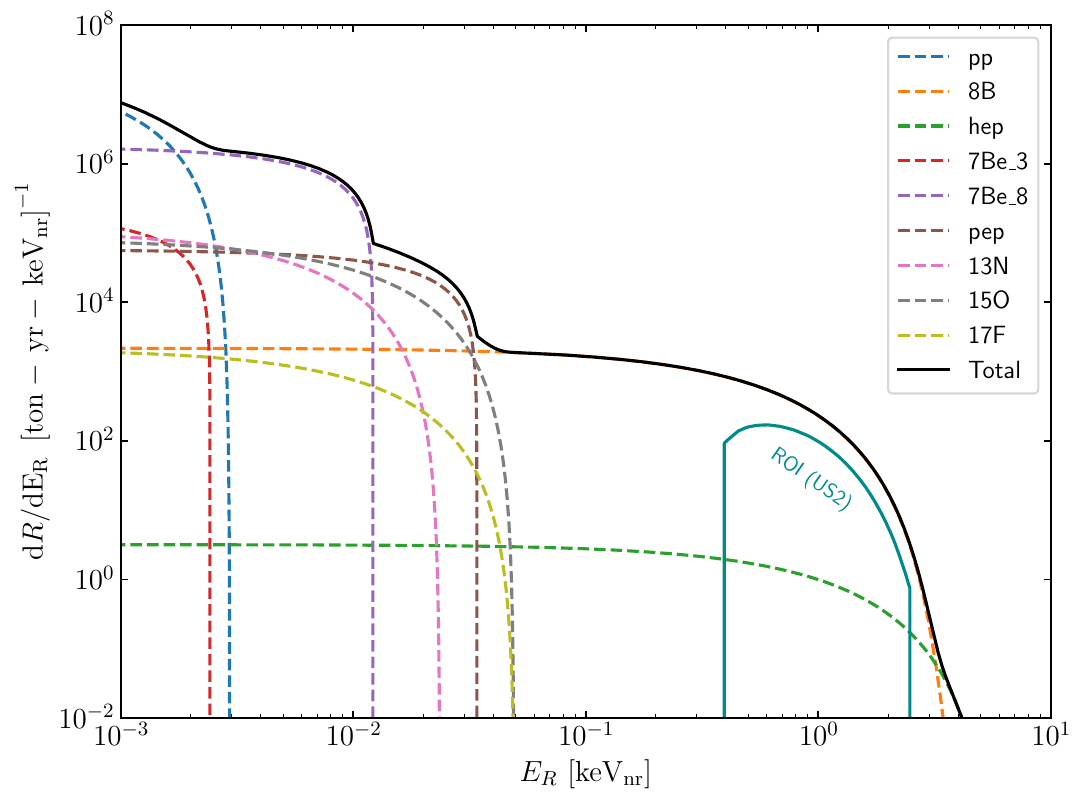}
    \includegraphics[scale=0.50]{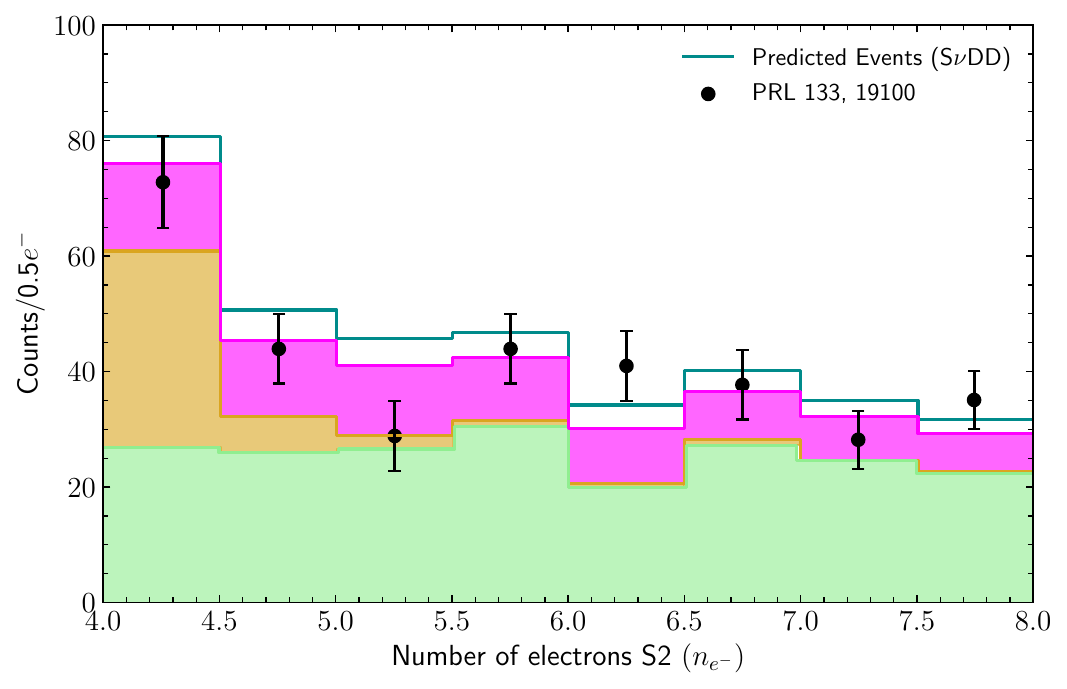}
    \caption{\justifying \textit{Left panel:} The black solid and coloured dashed lines correspond to the neutrino event rates obtained with S$\nu$DD \cite{Amaral:2023tbs}. The solid dark cyan line corresponds to the S2-only ROI of PandaX-4T.}
    \label{fig:rate-roi}
\end{figure}

The differential event rate as a function of $n_{e^{-}}$ is then expressed through a change of variables

\begin{equation}
    \frac{dR}{dn_{e^{-}}} =  \frac{dR}{dE_{R}} \frac{dE_{R}}{dn_{e^{-}}}
\end{equation} 
and the events per bin are
\begin{equation}\label{Ri}
  R_{i}
  = \,\mathcal{E}\,
    \int_{i}
      \frac{dR}{dn_{e^{-}}}\,dn_{e^{-}},
\end{equation}

where the integral is performed over the size of each bin $i$. We consider $\mathcal{E}$= 1.04 [ton $\cdot$ year] and 8 bins in the range $[4, 8]$ $n_{e^{-}}$. Our results are presented in Fig. \ref{fig:rate-roi}. In the left panel, we show the differential recoil rate induced by different sources of solar neutrinos, together with the total flux, and highlight the ROI of PandaX-4T, mainly sensitive to 8B neutrinos, with a small contribution from hep neutrinos. On the right panel of the Figure, we show the count rate at PandaX-4T S2-only search, along with their background data (with same color code as indicated in \ref{fig:counts-PandaX-DMe}. As can be noticed in the plot, we do not reproduce exactly the CE$\nu$NS rate predicted by PandaX-4T. While interesting, we restrict ourselves to using the PandaX-4T CE$\nu$NS expected background data in our analysis, but point out a sizable difference in our theoretical expectations in some energy bins.

\section{Details on the limit-setting}

Ws derived upper limits from a Poissonian binned likelihood with $\chi_{i}^{2}$-function in each bin $i$

\begin{align}
    \chi_{i}^{2}(\bar{\sigma}_{e} , m_{\chi})  =  2 \Big{[} N_{i}(\bar{\sigma}_{e} , m_{\chi}) + N_{i}^{bck} 
    &- N^{obs}_{i}  \ln \left( N_{i}(\bar{\sigma}_{e} , m_{\chi}) + N_{i}^{bck} \right) + \ln (N_{i}^{obs}!) \Big{]} \notag
\end{align}

and with total $\chi^{2}$ given by

\begin{equation}
    \chi^{2}(\bar{\sigma}_{e} , m_{\chi}) = \sum_{i} \chi_{i}^{2}(\bar{\sigma}_{e} , m_{\chi})
\end{equation}
Then the test statistic is given by

\begin{equation}
    \Delta \chi^{2} = \chi^{2}(\bar{\sigma}_{e} , m_{\chi}) - \chi^{2}_{min}
    = \chi^{2}_{\sigma}
\end{equation}

\begin{figure}
    \centering
    \includegraphics[scale=0.62]{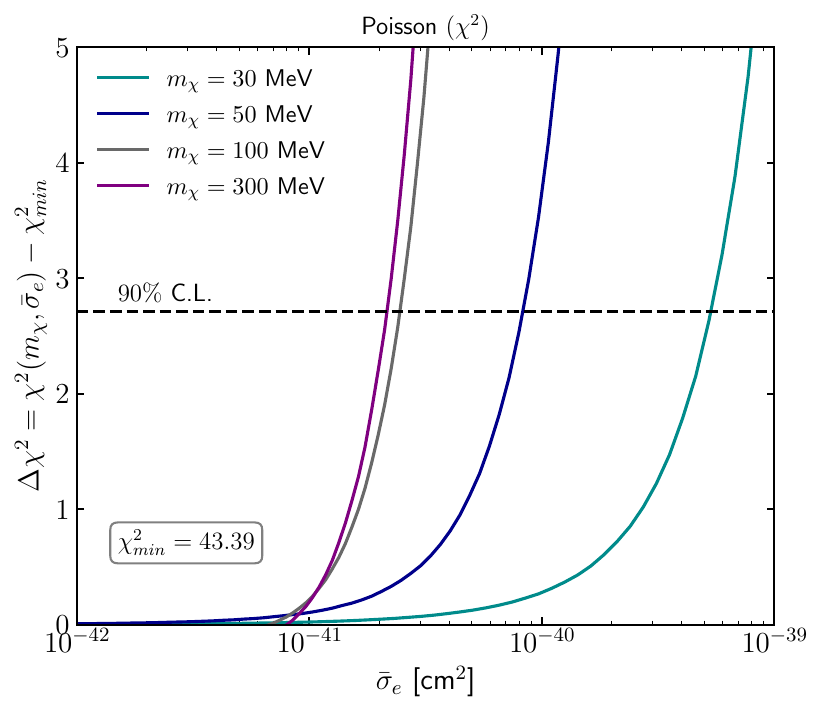}
    \includegraphics[scale=0.62]{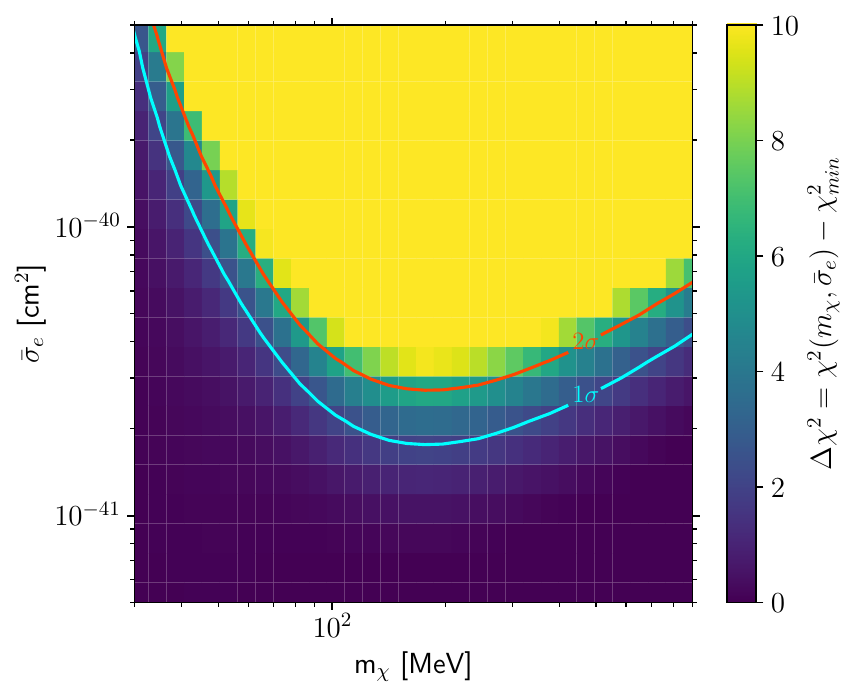}
    \caption{\justifying \textit{Left panel:} $ \Delta \chi^{2} = \chi^{2} - \chi^{2}_{min}$ as a function of $\bar{\sigma}_{e}$ for different values of $m_{\chi}$. \textit{Right panel:} Contour map on $\Delta \chi^2$ in the  $m_{\chi}-\bar{\sigma}_e$ plane.}
    \label{fig:chi2-plot}
\end{figure}

In the left panel of Figure \ref{fig:chi2-plot} we plot the $\Delta \chi^{2}$ as a function of $\bar{\sigma}_{e}$ for various dark matter masses, confronted with the 90\% C.L upper limit. In the right panel of the Figure, we show the upper limits obtained at 1$\sigma$ and 2$\sigma$, together with the contour map for different values of $\Delta \chi^2$

\end{document}